\documentclass[onecolumn]{revtex4}

\usepackage{graphicx}% Include figure files
\usepackage{dcolumn}% Align table columns on decimal point
\usepackage{bm}% bold math

\input epsf

\begin{document}

\title{\Large A Study of Generalized Second Law of Thermodynamics in
Magnetic Universe in the light of Non-Linear Electrodynamics}

\author{\bf~Tanwi~Bandyopadhyay$^1$\footnote{tanwib@gmail.com}~and~Ujjal~Debnath$^2$\footnote{ujjal@iucaa.ernet.in}}

\affiliation{$^1$Department of Mathematics,~Shri~Shikshayatan
College, 11, Lord~Sinha Road,~Kolkata-71, India.\\ $^2$Department
of Mathematics,~Bengal Engineering and Science
University,~Shibpur,~Howrah-711103, India.}

\begin{abstract}
In this work, we have considered the magnetic universe in
non-linear electrodynamics. The Einstein's field equations for
non-flat FRW model have been considered when the universe is
filled with the matter and magnetic field only. We have discussed
the validity of the generalized second law of thermodynamics of
the magntic universe bounded by Hubble, apparent, particle and
event horizons using Gibb's law and the first law of
thermodynamics for interacting and non-interacting scenarios. It
has been shown that the GSL is always satisfied for Hubble,
apparent and particle horizons but for event horizon, the GSL is
violated initially and satisfied at late stage of the universe.\\
\end{abstract}

\maketitle

\section{\normalsize\bf{Introduction}}

In present years the standard cosmological model based on
Friedmann-Robertson-Walker (FRW) with Maxwell electrodynamics has
got much attention and many interesting results are obtained. This
leads to a cosmological singularity at a finite time in the past
and result the energy density and curvature arbitrary large in the
very early epoch [1]. This singularity breaks the laws of physics
with mathematical inconsistency and physical incompleteness of any
cosmological model. There are some proposals to handle this
primordial singularity such as cosmological constant [2], non
minimal couplings [3], modifications of geometric structure of
space-time [4], non-equilibrium thermodynamics [5], Born-Infeld
type nonlinear electromagnetic field [6] and so on. Very strong
electromagnetic fields might help avoiding the occurrence of
space-time singularities in the cosmological context [7]. The
impact of very strong electromagnetic fields regarding the
causality issue in cosmology is also of relevance [8].\\

Studying the equations of the non-linear electrodynamics (NLED) is
an attractive subject of research in general relativity thanks to
the fact that such quantum phenomena as vacuum polarization can be
implemented in a classical model through their impact on the
properties of the background space-time. Recently a new approach
[9] has been taken to avoid the cosmic singularity through a
nonlinear extension of the Maxwell electromagnetic theory. The
associated Lagrangian and the resulting electrodynamics can
theoretically be justified based on different arguments. Exact
solutions of the Einstein's field equations coupled to NLED may
hint at the relevance of the non-linear effects in strong
gravitational and magnetic fields. For example, the nonlinear
terms can be added to the standard Maxwell Lagrangian by imposing
the existence of symmetries such as parity conservation, gauge
invariance, Lorentz invariance etc [10] as well as by the
introduction of first-order quantum corrections invariance to the
Maxwell electrodynamics [11]. Another interesting feature can be
viewed that an exact regular black hole solution has been recently
obtained proposing Einstein-dual nonlinear electrodynamics [12].
Also the General Relativity (GR) coupled with NLED effects can
explain the primordial inflation.\\

Since the discovery of black hole thermodynamics in 1970's,
physicists have been speculating that there should be some
relation between black hole thermodynamics and Einstein equations.
In Einstein gravity, the evidence of this connection was first
discovered in [13] by deriving the Einstein equation from the
proportionality of entropy and horizon area together with the
first law of thermodynamics $\delta Q = T dS$ in the Rindler
spacetime. He assumed that this relation holds for all Rindler
causal horizons through each space time point with $\delta Q$ and
$T$ interpreted as the energy flux and temperature seen by an
accelerated observer just inside the horizon. The horizon area
(geometric quantity) of black hole is associated with its entropy
(thermodynamical quantity), the surface gravity (geometric
quantity) is related with its temperature (thermodynamical
quantity) in black hole thermodynamics [14]. Verlinde [15] found
that the Friedmann equation in a radiation dominated
Friedmann-Robertson-Walker (FRW) universe can be written in an
analogous form of the Cardy-Verlinde formula, an entropy formula
for a conformal field theory. The thermodynamics in de Sitter
space–time was first investigated by Gibbons and Hawking in
[16].\\

The identity between Einstein equations and thermodynamical laws
has been applied in the cosmological context considering universe
as a thermodynamical system bounded by the apparent horizon. The
first law of thermodynamics for the cosmological horizon is given
by $-dE=TdS$, where $T=\frac{1}{2\pi l}$ is the Hawking
temperature, and $S=\frac{A}{4G}$ is the entropy with $A=4\pi
l^{2}$ and $G$ as the cosmological horizon area and Newton
constant respectively [17]. At the apparent horizon, the first law
of thermodynamics (on the apparent horizon) is shown to be
equivalent to Friedmann equations and the generalized second law
of thermodynamics (GSLT) is obeyed at the horizon. In a spatially
flat de Sitter space–time, the event horizon and the apparent
horizon of the Universe coincide and there is only one
cosmological horizon. In the usual standard big bang model a
cosmological event horizon does not exist. But for the
accelerating universe dominated by dark energy, the cosmological
event horizon separates from that of the apparent horizon. When
the apparent horizon and the event horizon of the Universe are
different, it was found that the first law and generalized second
law (GSL) of thermodynamics hold on the apparent horizon, while
they break down if one considers the event horizon [18]. On the
basis of the well known correspondence between the Friedmann
equation and the first law of thermodynamics of the apparent
horizon, Gong et al [19] argued that the apparent
horizon is the physical horizon in dealing with thermodynamics problems.\\

There are several studies in thermodynamics for dark energy filled
universe on apparent and event horizons [20]. Setare and Shafei
[21] showed that for the apparent horizon the first law is roughly
respected for different epochs while the second law of
thermodynamics is respected. Considering the interacting
holographic model of dark energy to investigate the validity of
the GSL of thermodynamics in a non-flat (closed) universe enclosed
by the event horizon, Setare [22] found that generalized second
law is respected for the special range of the deceleration
parameter. The transition from quintessence to phantom dominated
universe was considered and the conditions of the validity of GSL
in transition was studied in [23]. In the reference [24], a
Chaplygin gas dominated was considered and the GSL was
investigated taking into account the existence of the observer's
event horizon in accelerated universes and it was concluded that
for the initial stage of Chaplygin gas dominated expansion, the
GSL of gravitational thermodynamics is fulfilled. Recently, the
GSL of thermodynamics on Hubble, apparent, particle and event
horizons have been extensively studied in [25].\\

In this work, we have briefly discussed the Maxwell's
electrodynamics in linear and non-linear forms in section II. The
energy density and pressure for non-linear electrodynamics have
been written in magnetic universe only. The Einstein's field
equations for non-flat FRW model have been considered if the
universe is filled with the matter and magnetic field only. The
interaction between matter and magnetic field have been considered
in section III and some particular form of interaction term, we
have found the solutions of magnetic field and the energy density
of matter. In section IV, the validity of generalized second law
of thermodynamics have been investigated on Hubble, apparent,
particle and event horizons using Gibb's law and the first law of
thermodynamics. Finally, we have made some concluding remarks.\\

\section{\normalsize\bf{Brief overview of Non-linear Electrodynamics}}

The Lagrangian density in Maxwell's electrodynamics can be written
as [26]

\begin{equation}
{\cal
L}=-\frac{1}{4\mu_{0}}~F^{\mu\nu}F_{\mu\nu}=-\frac{1}{4\mu_{0}}~F
\end{equation}

where $F^{\mu\nu}$ is the electromagnetic field strength tensor
and $\mu_{0}$ is the magnetic permeability. The canonical
energy-momentum tensor is then given by

\begin{equation}
T_{\mu\nu}=\frac{1}{\mu_{0}}\left(F_{\mu\alpha}F^{\alpha}_{\nu}+\frac{1}{4}~Fg_{\mu\nu}
\right)
\end{equation}

The homogeneous, isotropic FRW model of the universe is given by

\begin{equation}
ds^{2}=dt^{2}-a^{2}(t)[\frac{dr^{2}}{1-kr^{2}}+r^{2}(d\theta^{2}+sin^{2}\theta
d\phi^{2})]
\end{equation}

Since the spatial section of FRW geometry are isotropic,
electromagnetic fields can generate such a universe only if an
averaging procedure is performed [27]. Applying standard spatial
averaging process for electric field $E_{i}$ and magnetic field
$B_{i}$, set

\begin{equation}
<E_{i}>=0,~~<B_{i}>=0,~~<E_{i}E_{j}>=-\frac{1}{3}~E^{2}g_{ij},
~~<B_{i}B_{j}>=-\frac{1}{3}~B^{2}g_{ij},~~<E_{i}B_{j}>=0.
\end{equation}

So from equation (4) we get,

\begin{equation}
<F_{\mu\alpha}F^{\alpha}_{\nu}>=\frac{2}{3}\left(\epsilon_{0}E^{2}+
\frac{B^{2}}{\mu_{0}}
\right)u_{\mu}u_{\nu}+\frac{1}{3}\left(\epsilon_{0}E^{2}-
\frac{B^{2}}{\mu_{0}} \right)g_{\mu\nu}
\end{equation}

where $u_{\mu}$ is the fluid 4-velocity. Now comparing with the
average value of energy momentum tensor

\begin{equation}
<T_{\mu\nu}>=(\rho+p)u_{\mu}u_{\nu}-pg_{\mu\nu}
\end{equation}

the energy density and pressure have the forms

\begin{equation}
\rho=\frac{1}{2}\left(\epsilon_{0}E^{2}+ \frac{B^{2}}{\mu_{0}}
\right)~,~~~~p=\frac{1}{3}~\rho
\end{equation}

This shows that the Maxwell's electrodynamics generates only the
radiation type fluid in FRW universe.\\

Here we consider the generalization of Maxwell electro-magnetic
Lagrangian up to the second order terms of the fields as [26]

\begin{equation}
{\cal L}=-\frac{1}{4\mu_{0}}~F+\omega F^{2}+\eta F^{*2}
\end{equation}

where $\omega$ and $\eta$ are arbitrary constants,

\begin{equation}
F^{*}\equiv F^{*}_{\mu\nu}F^{\mu\nu}
\end{equation}

and $F^{*}_{\mu\nu}$ is the dual of $F_{\mu\nu}$. So the
corresponding energy-momentum tensor for non-linear
electro-magnetic theory has the form

\begin{equation}
T_{\mu\nu}=-4~\frac{\partial {\cal L}}{\partial
F}~F^{\alpha}_{\mu}F_{\alpha\nu}+\left(\frac{\partial {\cal
L}}{\partial F^{*}}~F^{*}-{\cal L}\right)g_{\mu\nu}
\end{equation}

Now we consider the homogeneous electric field $E$ in plasma gives
rise to an electric current of charged particles and then rapidly
decays. So the squared magnetic field $B^{2}$ dominates over
$E^{2}$, i.e., in this case, the average value $<E^{2}>\approx 0$
and hence $F=2B^{2}$. So $F$ is now only the function of magnetic
field (vanishing electric component) and hence the FRW universe
may be called the {\it magnetic universe}. Now, similar to above
discussions, we get the energy density and the pressure for
magnetic field have the forms [26]

\begin{equation}
\rho_{B}=\frac{B^{2}}{2\mu_{0}}(1-8\mu_{0}\omega B^{2})
\end{equation}

and

\begin{equation}
p_{B}=\frac{B^{2}}{6\mu_{0}}(1-40\mu_{0}\omega
B^{2})=\frac{1}{3}\rho_{B}-\frac{16}{3}\omega B^{4}
\end{equation}

It is to be noted that the density of the magnetic field must be
positive, so the magnetic field $B$ must be satisfied
$B<\frac{1}{2\sqrt{2\mu_{0}\omega}}$. Comparing (7) and (12), we say
that $\frac{16}{3}\omega B^{4}$ is the correction term of the EOS for
radiation in the generalization of Maxwell's electrodynamics.\\

The Einstein's field equations are given by

\begin{equation}
H^{2}+\frac{k}{a^{2}}=\frac{8\pi G}{3}\rho_{total}
\end{equation}

and

\begin{equation}
\dot{H}-\frac{k}{a^{2}}=-4\pi G(\rho_{total}+p_{total})
\end{equation}

where

\begin{equation}
\rho_{total}=\rho_{m}+\rho_{B}=\rho_{m}+\frac{B^{2}}{2\mu_{0}}(1-8\mu_{0}\omega
B^{2})
\end{equation}

and

\begin{equation}
p_{total}=p_{m}+p_{B}=w_{m}\rho_{m}+\left(\frac{1}{3}\rho_{B}-\frac{16}{3}\omega
B^{4}\right)
\end{equation}

where, $\rho_{m}$ and $p_{m}$ are energy density and pressure for
matter obeys the equation of state $p_{m}=w_{m}\rho_{m}$.\\

Now the energy-conservation equation is

\begin{equation}
\dot{\rho}_{total}+3H(\rho_{total}+p_{total})=0
\end{equation}

where $H=\frac{\dot{a}}{a}$ is the Hubble parameter.\\

\section{\normalsize\bf{Interaction between matter and Magnetic field}}

Here we consider the interaction between matter and magnetic
field. So the conservation equation (17) becomes

\begin{equation}
\dot{\rho}_{m}+3H(1+w_{m})\rho_{m}=Q
\end{equation}

and

\begin{equation}
\dot{\rho}_{B}+3H(\rho_{B}+p_{B})=-Q
\end{equation}

For simplicity of the calculation, we take the interaction
component as

\begin{equation}
Q=3\delta\frac{B}{\mu_{0}}(1-16\mu_{0}\omega B^{2})H
\end{equation}

where $\delta$ is a small positive quantity. \\

Using the expressions of $\rho_{E}$, $p_{E}$, $\rho_{m}$ and
$p_{m}$, the above two equations can be solved to obtain

\begin{equation}
B=-\frac{3}{2}\delta+\frac{B_{0}}{a^{2}},~~~~~B_{0}~\text{being a
constant}
\end{equation}

and

\begin{equation}
\rho_{m}=\frac{3\delta}{2\mu_{0}}\left[-\frac{32B_{0}^{2}\mu_{0}\omega}{3(w_{m}-1)a^{6}}
+\frac{144B_{0}^{2}\delta\mu_{0}\omega}{(3w_{m}-1)a^{4}}-\frac{\delta(1-36\delta^{2}\mu_{0}\omega)}{w_{m}+1}
-\frac{2B_{0}(108\delta^{2}\mu_{0}\omega-1)}{(3w_{m}+1)a^{2}}\right]+\rho_{0}a^{-3(1+w_{m})}
\end{equation}

where $\rho_{0}$ is an integration constant. For the above solutions, we can verified that the interaction term $Q$
always decays with the evolution of the universe.\\

\section{\normalsize\bf{Generalized Second Law of Thermodynamics}}

In this section, the validity of the generalized second law of
thermodynamics is studied. It states that, the sum of entropy of
total matter enclosed by the horizon and the entropy of the
horizon does not decrease with time. In the following, we consider
Hubble, apparent, particle and event horizons. The variation of
entropy inside the horizon will be calculated via Gibb's equation
and the variation of entropy on the horizon will be calculated
using first law of thermodynamics. Hence we shall examine the
validity of GSL of thermodynamics of the universe bounded by
the above mentioned horizons.\\

\subsection{\normalsize\bf{Hubble Horizon}}

We know that radius of Hubble horizon [25],

\begin{equation}
R_{H}=\frac{1}{H}
\end{equation}

Therefore

\begin{equation}
\dot{R}_{H}=-\frac{\dot{H}}{H^{2}}=\frac{\frac{2B^{2}}{\mu_{0}}(1-16\mu_{0}\omega
B^{2})+3(1+w_{m})\rho_{m}-\frac{3k}{4\pi Ga^{2}}}
{\frac{B^{2}}{\mu_{0}}(1-8\mu_{0}\omega
B^{2})+2\rho_{m}-\frac{3k}{4\pi Ga^{2}}}
\end{equation}

Considering the net amount of energy crossing through the Hubble
horizon in time $dt$ as [28]

\begin{equation}
-dE=4\pi R_{H}^{3}H(\rho_{total}+p_{total})dt
\end{equation}

and assuming the validity of first law of thermodynamics on the
Hubble horizon, i.e,

\begin{equation}
-dE=T_{H}dS_{H}
\end{equation}

we have

\begin{equation}
\frac{dS_{H}}{dt}=\frac{4\pi
R_{H}^{3}H}{T_{H}}\left[\frac{2B^{2}}{3\mu_{0}}(1-16\mu_{0}\omega
B^{2})+(1+w_{m})\rho_{m}\right]
\end{equation}

Again from the Gibbs' eqn [18, 24]

\begin{equation}
T_{H}dS_{I}=dE_{I}+p_{total}dV
\end{equation}

we have

\begin{equation}
\frac{dS_{I}}{dt}=\frac{4\pi
R_{H}^{2}}{T_{H}}\left[\frac{2B^{2}}{3\mu_{0}}(1-16\mu_{0}\omega
B^{2})+(1+w_{m})\rho_{m}\right](\dot{R}_{H}-1)
\end{equation}

where $S_{I}$ is the entropy inside the horizon, $V=\frac{4}{3}\pi
R^{3}$ is the volume of the horizon universe,
$E_{I}=\rho_{total}V$ is the internal energy and $T$
stands for the Hawking temperature.\\

From eqns (27) and (29), the rate of change of the total entropy
becomes

\begin{equation}
\frac{d}{dt}(S_{H}+S_{I})=\frac{4\pi
R_{H}^{2}}{T_{H}}\left[\frac{2B^{2}}{3\mu_{0}}(1-16\mu_{0}\omega
B^{2})+(1+w_{m})\rho_{m}\right]\left[\frac{\frac{2B^{2}}{\mu_{0}}(1-16\mu_{0}\omega
B^{2})+3(1+w_{m})\rho_{m}-\frac{3k}{4\pi Ga^{2}}}
{\frac{B^{2}}{\mu_{0}}(1-8\mu_{0}\omega
B^{2})+2\rho_{m}-\frac{3k}{4\pi Ga^{2}}}\right]
\end{equation}

Here the expressions of $B$ and $\rho_{m}$ are given in (21) and
(22). We plot the rate of change of total entropy of Hubble
horizon i.e., $\dot{S}_{H}+\dot{S}_{I}$ against redshift $z$ in
figures 1 and 2, without ($\delta=0$) and with interaction
($\delta=0.001$) respectively for different matter components
i.e., $w_{m}=1/3$ (solid line), $w_{m}=0$ (dotted line) and
$w_{m}=-0.5$ (dashed line) and $k=0$ (red line), $k=+1$ (green
line) and $k=-1$ (blue line). From these figures, we conclude that
the GSL is always valid for Hubble horizon for non-interacting
and interacting scenarios of the magnetic universe.\\

\begin{figure}
\includegraphics[height=2.0in]{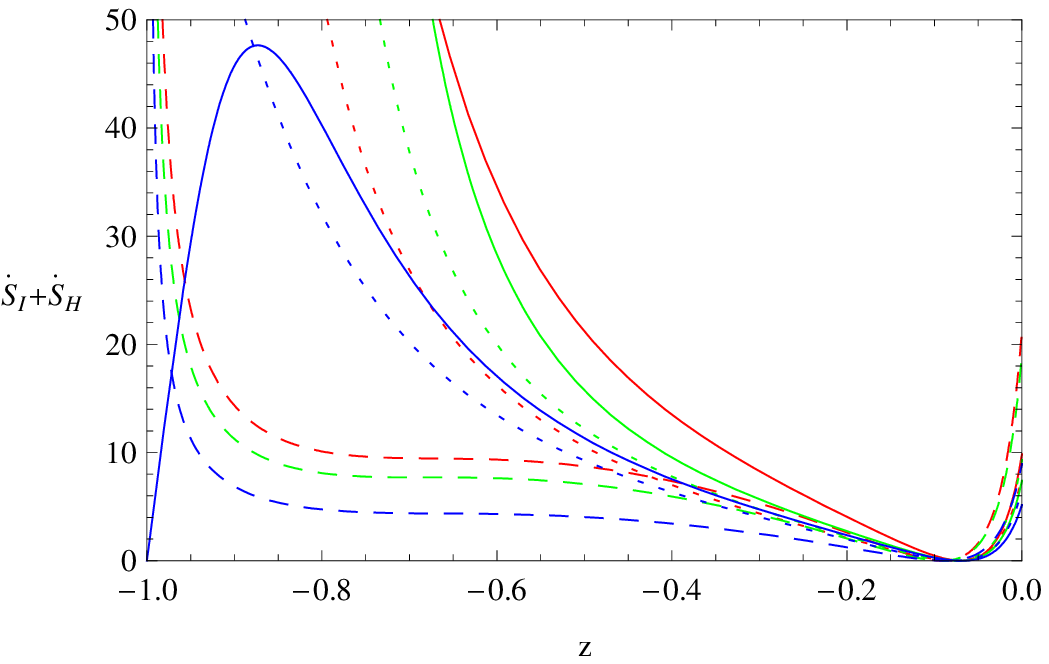}~~~~~~
\includegraphics[height=2.0in]{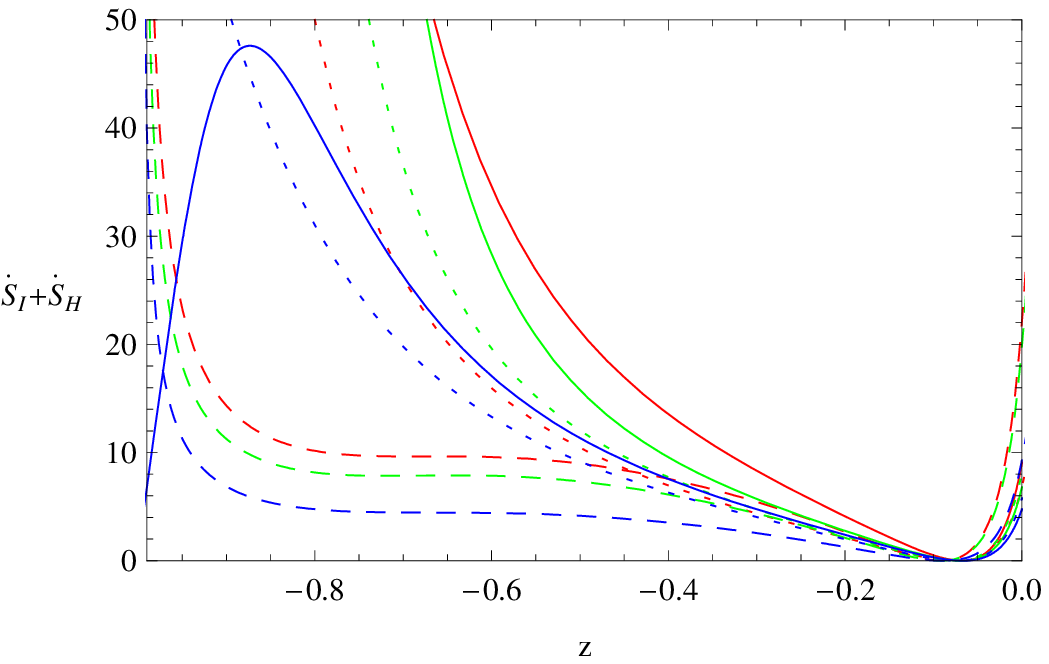}\\
\vspace{1mm}
~~~~~Fig.1~~~~~~~~~~~~~~~~~~~~~~~~~~~~~~~~~~~~~~~~~~~~~~~~~~~~~~~~~~~~~~~~~~~~~~~~~~~~~Fig.2\\

\vspace{6mm} ~~~~~~~~~~~~Figs.1 and 2 represent rate of change of
total entropy of Hubble horizon i.e., $\dot{S}_{H}+\dot{S}_{I}$
against redshift $z$ without and with interaction respectively for
$w_{m}=1/3$ (solid line), $w_{m}=0$ (dotted line)
and $w_{m}=-0.5$ (dashed line) and $k=0$ (red line), $k=+1$ (green line) and $k=-1$ (blue line).\\

\vspace{6mm}

\end{figure}

\subsection{\normalsize\bf{Apparent Horizon}}

We know that radius of apparent horizon [25],

\begin{equation}
R_{A}=\frac{1}{\sqrt{H^{2}+\frac{k}{a^{2}}}}
\end{equation}

Therefore

\begin{equation}
\dot{R}_{A}=-\frac{H(\dot{H}-\frac{k}{a^{2}})}{(H^{2}+\frac{k}{a^{2}})^{3/2}}
=\frac{\left[\frac{B^{2}}{\mu_{0}}(1-8\mu_{0}\omega
B^{2})+2\rho_{m}-\frac{3k}{4\pi
Ga^{2}}\right]^{1/2}\left[\frac{2B^{2}}{\mu_{0}}(1-16\mu_{0}\omega
B^{2})+3(1+w_{m})\rho_{m}\right]}
{\left[\frac{B^{2}}{\mu_{0}}(1-8\mu_{0}\omega
B^{2})+2\rho_{m}\right]^{3/2}}
\end{equation}

Considering the net amount of energy crossing through the apparent
horizon in time $dt$ as [28]

\begin{equation}
-dE=4\pi R_{A}^{3}H(\rho_{total}+p_{total})dt
\end{equation}

and assuming the validity of first law of thermodynamics on the
apparent horizon, i.e,

\begin{equation}
-dE=T_{A}dS_{A}
\end{equation}

we have

\begin{equation}
\frac{dS_{A}}{dt}=\frac{4\pi
R_{A}^{3}H}{T_{A}}\left[\frac{2B^{2}}{3\mu_{0}}(1-16\mu_{0}\omega
B^{2})+(1+w_{m})\rho_{m}\right]
\end{equation}

Again from the Gibbs' eqn [18, 24]

\begin{equation}
T_{A}dS_{I}=dE_{I}+p_{total}dV
\end{equation}

we have

\begin{equation}
\frac{dS_{I}}{dt}=\frac{4\pi
R_{A}^{2}}{T_{A}}\left[\frac{2B^{2}}{3\mu_{0}}(1-16\mu_{0}\omega
B^{2})+(1+w_{m})\rho_{m}\right](\dot{R}_{A}-HR_{A})
\end{equation}

From eqns (35) and (37), the rate of change of the total entropy
becomes

\begin{equation}
\frac{d}{dt}(S_{A}+S_{I})=\frac{12\pi
R_{A}^{2}}{T_{A}}\left[\frac{2B^{2}}{3\mu_{0}}(1-16\mu_{0}\omega
B^{2})+(1+w_{m})\rho_{m}\right]^{2}
\frac{\left[\frac{B^{2}}{\mu_{0}}(1-8\mu_{0}\omega
B^{2})+2\rho_{m}-\frac{3k}{4\pi
Ga^{2}}\right]^{1/2}}{\left[\frac{B^{2}}{\mu_{0}}(1-8\mu_{0}\omega
B^{2})+2\rho_{m}\right]^{3/2}}
\end{equation}

Here the expressions of $\rho_{m}$ and $B$ are given in (21) and
(22). We plot the rate of change of total entropy of apparent
horizon i.e., $\dot{S}_{A}+\dot{S}_{I}$ against redshift $z$ in
figures 3 and 4, without ($\delta=0$) and with interaction
($\delta=0.001$) respectively for different matter components
i.e., $w_{m}=1/3$ (solid line), $w_{m}=0$ (dotted line) and
$w_{m}=-0.5$ (dashed line) and $k=0$ (red line), $k=+1$ (green
line) and $k=-1$ (blue line). From these figures, we conclude that
the GSL is always valid for apparent horizon for non-interacting
and interacting scenarios of the magnetic universe.\\

\begin{figure}
\includegraphics[height=2.0in]{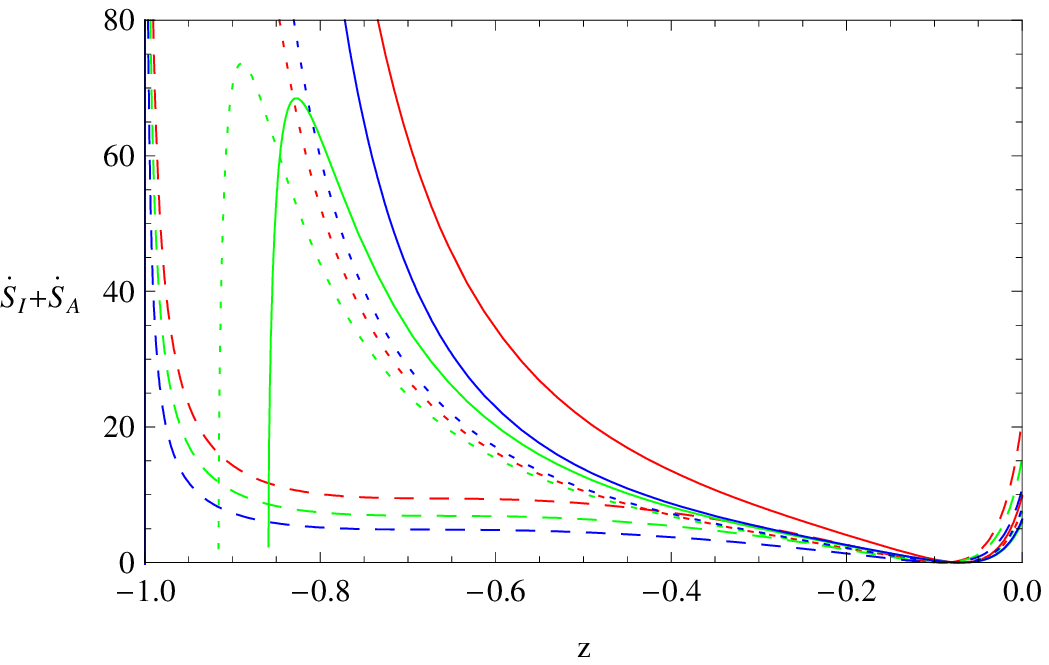}~~~~~~
\includegraphics[height=2.0in]{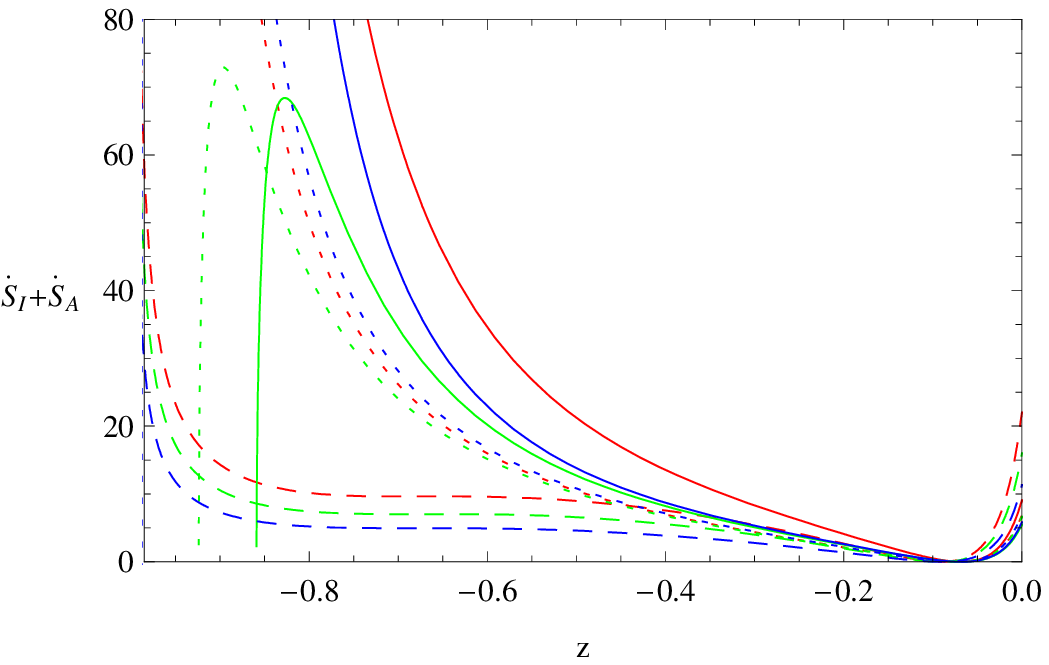}\\
\vspace{1mm}
~~~~~Fig.3~~~~~~~~~~~~~~~~~~~~~~~~~~~~~~~~~~~~~~~~~~~~~~~~~~~~~~~~~~~~~~~~~~~~~~~~~~~~~~~~~Fig.4\\

\vspace{6mm} ~~~~~~~~~~~~Figs.3 and 4 represent rate of change of
total entropy of apparent horizon i.e., $\dot{S}_{A}+\dot{S}_{I}$
against redshift $z$ without and with interaction respectively for
$w_{m}=1/3$ (solid line), $w_{m}=0$ (dotted line)
and $w_{m}=-0.5$ (dashed line) and $k=0$ (red line), $k=+1$ (green line) and $k=-1$ (blue line).\\

\vspace{6mm}

\end{figure}

\subsection{\normalsize\bf{Particle Horizon}}

The horizon radius is given by [25]

\begin{equation}
R_{P}=a\int_{0}^{a}\frac{da}{Ha^{2}}
\end{equation}

The differential eqn of which can be written as

\begin{equation}
\dot{R}_{P}=HR_{P}+1
\end{equation}

Considering the net amount of energy crossing through the particle
horizon in time $dt$ as [28]

\begin{equation}
-dE=4\pi R_{P}^{3}H(\rho_{total}+p_{total})dt
\end{equation}

and assuming the validity of first law of thermodynamics on the
particle horizon, i.e,

\begin{equation}
-dE=T_{P}dS_{P}
\end{equation}

we have

\begin{equation}
\frac{dS_{P}}{dt}=\frac{4\pi
R_{P}^{3}H}{T_{P}}\left[\frac{2B^{2}}{3\mu_{0}}(1-16\mu_{0}\omega
B^{2})+(1+w_{m})\rho_{m}\right]
\end{equation}

Again from the Gibbs' eqn [18, 24]

\begin{equation}
T_{P}dS_{I}=dE_{I}+p_{total}dV
\end{equation}

we have

\begin{equation}
\frac{dS_{I}}{dt}=\frac{-4\pi
R_{P}^{2}}{T_{P}}\left[\frac{2B^{2}}{3\mu_{0}}(1-16\mu_{0}\omega
B^{2})+(1+w_{m})\rho_{m}\right]
\end{equation}

From eqns (43) and (45), the rate of change of the total entropy
becomes

\begin{equation}
\frac{d}{dt}(S_{P}+S_{I})=\frac{4\pi
R_{P}^{2}}{T_{P}}\left[\frac{2B^{2}}{3\mu_{0}}(1-16\mu_{0}\omega
B^{2})+(1+w_{m})\rho_{m}\right](HR_{P}+1)
\end{equation}

We plot the rate of change of total entropy of particle horizon
i.e., $\dot{S}_{P}+\dot{S}_{I}$ against redshift $z$ in figures 5
and 6, without ($\delta=0$) and with interaction ($\delta=0.001$)
respectively for different matter components i.e., $w_{m}=1/3$
(solid line), $w_{m}=0$ (dotted line) and $w_{m}=-0.5$ (dashed
line) and $k=0$ (red line), $k=+1$ (green line) and $k=-1$ (blue
line). From these figures, we conclude that the GSL is always
valid for particle horizon for non-interacting
and interacting scenarios of the magnetic universe.\\

\begin{figure}
\includegraphics[height=2.0in]{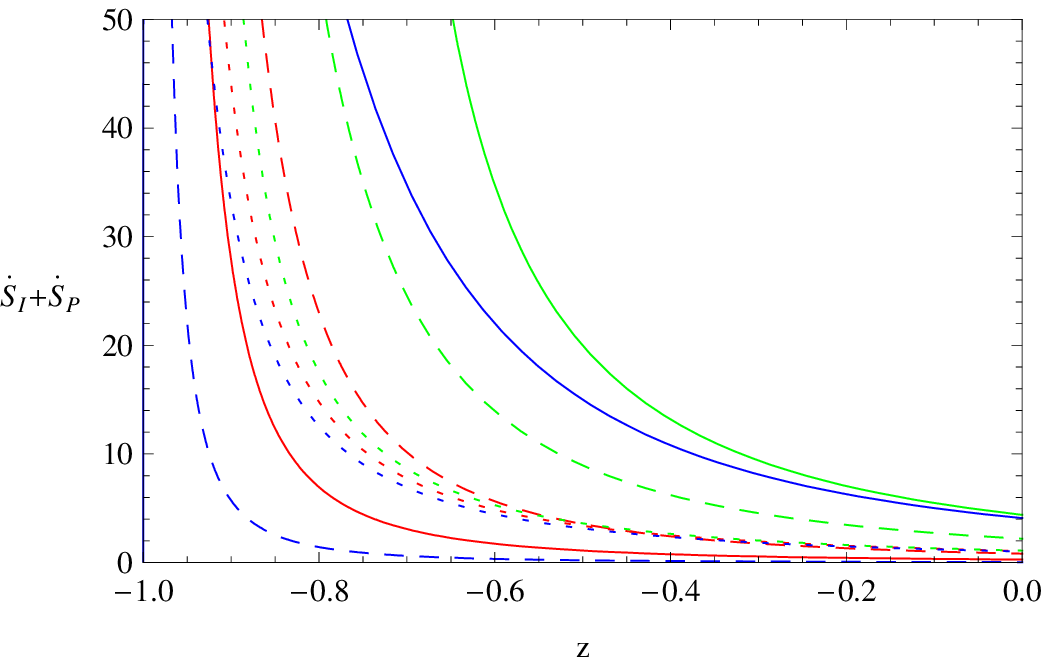}~~~~~~
\includegraphics[height=2.0in]{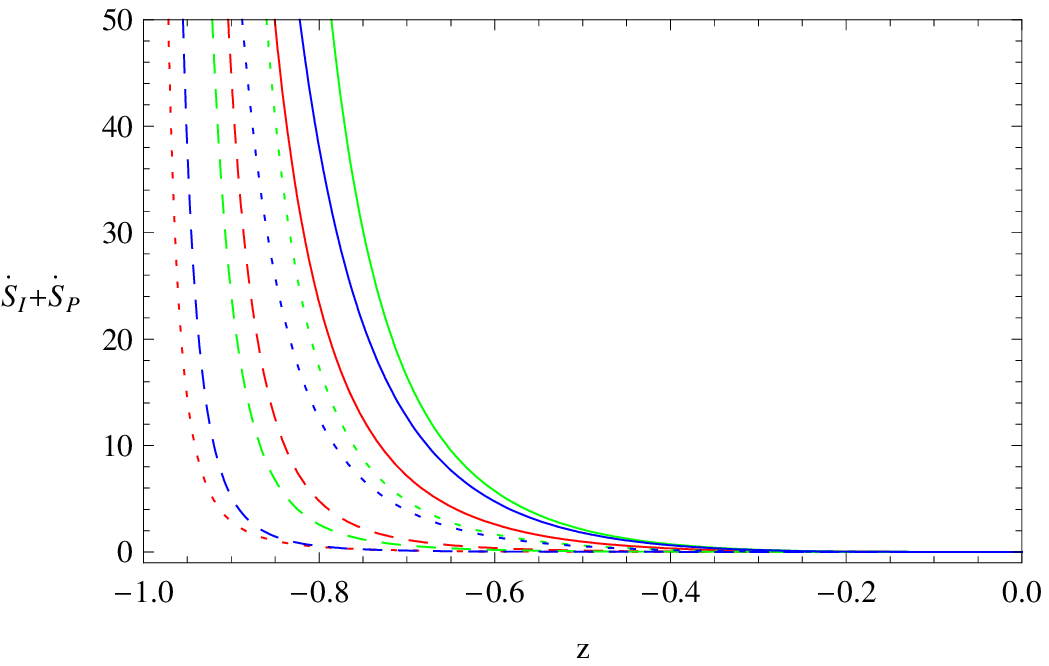}\\
\vspace{1mm}
~~~~~Fig.5~~~~~~~~~~~~~~~~~~~~~~~~~~~~~~~~~~~~~~~~~~~~~~~~~~~~~~~~~~~~~~~~~~~~~~~~~~~~~~~~~Fig.6\\

\vspace{6mm} ~~~~~~~~~~~~Figs.5 and 6 represent rate of change of
total entropy of particle horizon i.e., $\dot{S}_{P}+\dot{S}_{I}$
against redshift $z$ without and with interaction respectively for
$w_{m}=1/3$ (solid line), $w_{m}=0$ (dotted line)
and $w_{m}=-0.5$ (dashed line) and $k=0$ (red line), $k=+1$ (green line) and $k=-1$ (blue line).\\

\vspace{6mm}

\end{figure}

\subsection{\normalsize\bf{Event Horizon}}

The horizon radius is given by [25]

\begin{equation}
R_{E}=a\int_{a}^{\infty}\frac{da}{Ha^{2}}
\end{equation}

The differential eqn of which can be written as

\begin{equation}
\dot{R}_{E}=HR_{E}-1
\end{equation}

Considering the net amount of energy crossing through the event
horizon in time $dt$ as [28]

\begin{equation}
-dE=4\pi R_{E}^{3}H(\rho_{total}+p_{total})dt
\end{equation}

and assuming the validity of first law of thermodynamics on the
event horizon, i.e,

\begin{equation}
-dE=T_{E}dS_{E}
\end{equation}

we have

\begin{equation}
\frac{dS_{E}}{dt}=\frac{4\pi
R_{E}^{3}H}{T_{E}}\left[\frac{2B^{2}}{3\mu_{0}}(1-16\mu_{0}\omega
B^{2})+(1+w_{m})\rho_{m}\right]
\end{equation}

Again from the Gibbs' eqn [18, 24]

\begin{equation}
T_{E}dS_{I}=dE_{I}+p_{total}dV
\end{equation}

we have

\begin{equation}
\frac{dS_{I}}{dt}=\frac{-4\pi
R_{E}^{2}}{T_{E}}\left[\frac{2B^{2}}{3\mu_{0}}(1-16\mu_{0}\omega
B^{2})+(1+w_{m})\rho_{m}\right]
\end{equation}

From eqns (51) and (53), the rate of change of the total entropy
becomes

\begin{equation}
\frac{d}{dt}(S_{E}+S_{I})=\frac{4\pi
R_{E}^{2}}{T_{E}}\left[\frac{2B^{2}}{3\mu_{0}}(1-16\mu_{0}\omega
B^{2})+(1+w_{m})\rho_{m}\right](HR_{E}-1)
\end{equation}

We plot the rate of change of total entropy of event horizon i.e.,
$\dot{S}_{E}+\dot{S}_{I}$ against redshift $z$ in figures 7 and 8,
without ($\delta=0$) and with interaction ($\delta=0.001$)
respectively for different matter components i.e., $w_{m}=1/3$
(solid line), $w_{m}=0$ (dotted line) and $w_{m}=-0.5$ (dashed
line) and $k=0$ (red line), $k=+1$ (green line) and $k=-1$ (blue
line). From these figures, we see that the rate of change of total
entropy is negative level upto certain stage (about $z>-0.1$) and
positive level after that stage for non-interacting and
interacting scenarios of the magnetic universe bounded by event
horizon. So we conclude that GSL is not satisfied upto certain
stage and after this stage, it is always satisfied for magnetic
universe bounded by event horizon.\\

\begin{figure}
\includegraphics[height=2.0in]{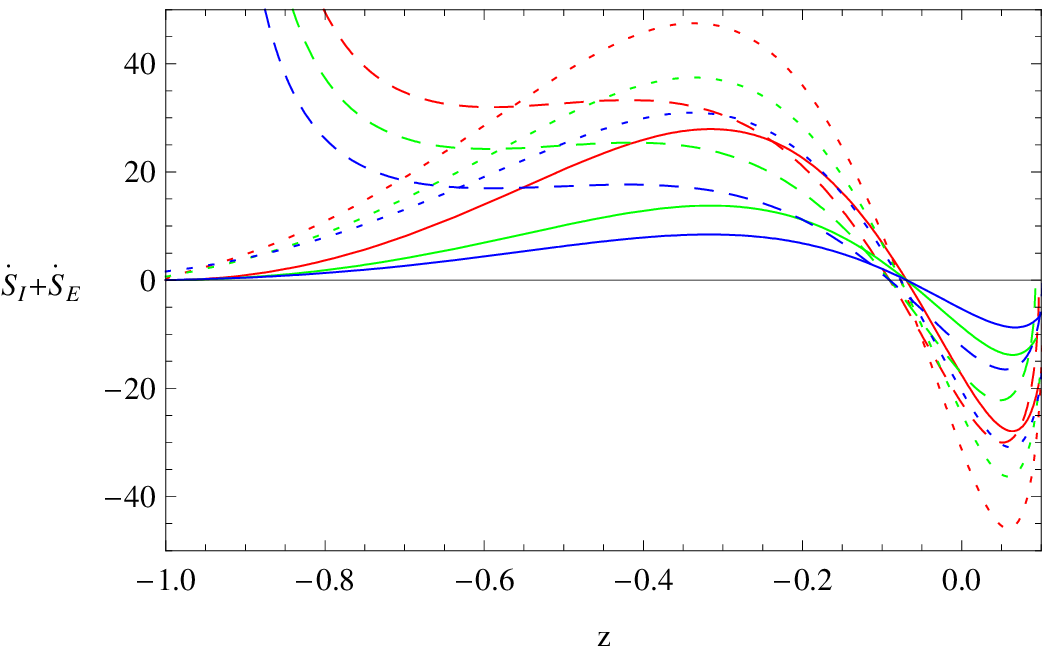}~~~~~~
\includegraphics[height=2.0in]{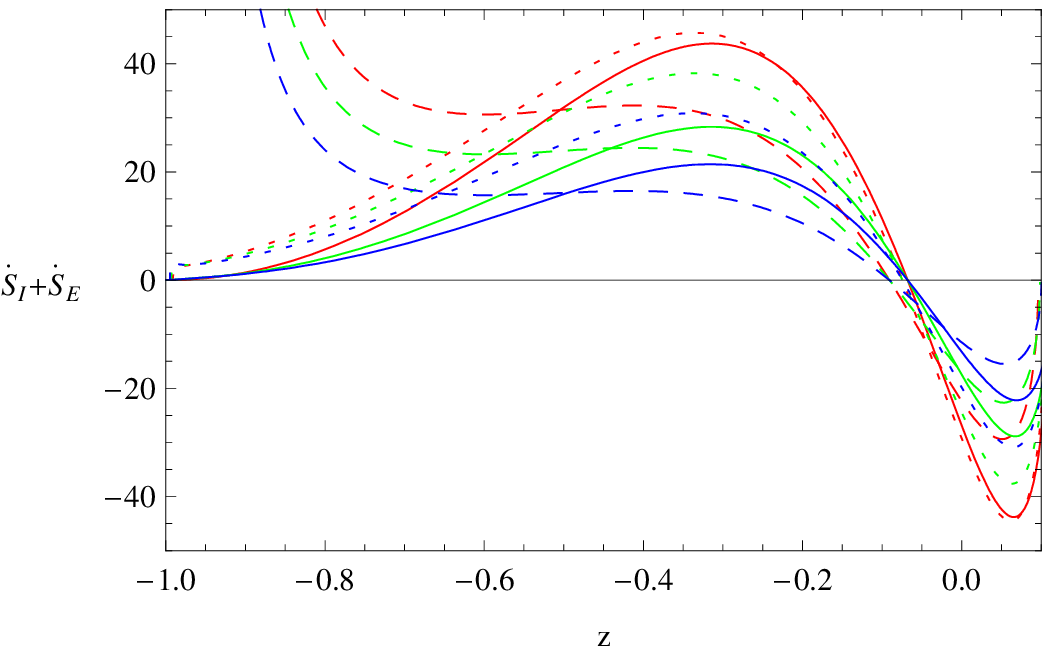}\\
\vspace{1mm}
~~~~~Fig.7~~~~~~~~~~~~~~~~~~~~~~~~~~~~~~~~~~~~~~~~~~~~~~~~~~~~~~~~~~~~~~~~~~~~~~~~~~~~~~~~~Fig.8\\

\vspace{6mm} ~~~~~~~~~~~~Figs.7 and 8 represent rate of change of
total entropy of event horizon i.e., $\dot{S}_{E}+\dot{S}_{I}$
against redshift $z$ without and with interaction respectively for
$w_{m}=1/3$ (solid line), $w_{m}=0$ (dotted line)
and $w_{m}=-0.5$ (dashed line) and $k=0$ (red line), $k=+1$ (green line) and $k=-1$ (blue line).\\

\vspace{6mm}

\end{figure}

\section{\normalsize\bf{Discussions}}

In this work, we have briefly discussed the Maxwell's
electrodynamics in linear and non-linear forms. The energy density
and pressure for non-linear electrodynamics have been written in
magnetic universe only. The Einstein's field equations for
non-flat FRW model have been considered when the universe is
filled with the matter and magnetic field only. The interaction
between matter and magnetic field have been considered and some
particular form of interaction term, we have found the solutions
of magnetic field and the energy density of matter. The
interaction term is always decreases with the time.\\

In the present work, our endeavor was to investigate the validity
of the generalized second law of thermodynamics of the magnetic
universe bounded by the Hubble, apparent, particle and event
horizons. It states that, the sum of entropy of total matter
enclosed by the horizon and the entropy of the horizon does not
decrease with time. The variation of entropy inside the horizon
have been calculated via Gibb's equation and the variation of
entropy on the horizon have been calculated using first law of
thermodynamics. We have investigated the generalized second law of
the universe bounded by the above mentioned horizons.\\

In figures 1 - 8, we have drawn the variation of total entropy on
Hubble, apparent, particle and event horizons against redshift $z$
for non-interacting ($\delta=0$) and interacting ($\delta=0.001$)
scenarios of magnetic universe for $k=0,\pm 1$ and
$w_{m}=0,1/3,-0.5$. From figures 1 - 6, we have seen that the rate
of change of total entropy is always positive level for redshift
$z$ decreases for interacting and non-interacting scenarios of the
magnetic universe bounded by Hubble, apparent and particle
horizons. But figures 7 - 8 show the rate of change of total
entropy is negative level upto certain stage (about $z>-0.1$) and
positive level after that stage for non-interacting and
interacting scenarios of the magnetic universe bounded by event
horizon. So these imply, the GSL is always satisfied for magnetic
universe bounded by Hubble, apparent and particle horizons. Also
GSL is not satisfied upto certain stage and after this stage, it
is always satisfied for magnetic universe bounded
by event horizon.\\

{\bf Acknowledgement}:\\

The authors are thankful to IUCAA, Pune, for their warm
hospitality and excellent research facilities where most of the
work has been done during a visit under the Associateship
Programme. Also TB wants to thank UGC for providing with a project
on thermodynamics.\\\\

{\bf References:}\\

[1] E. W. Kolb and M. S. Turner, \textit{Addison-Wesley, Redwood City, CA} (1990).\\\

[2] W. de Sitter,  \textit{Proc. K. Ned. Akad. Wet.}, \textbf{19} 1217 (1917).\\\

[3] M. Novello and J. M. Salim,  \textit{Phys. Rev. D} \textbf{20} 377 (1979).\\\

[4] M. Novello et al, \textit{IJMPA} \textbf{1} 641 (1993).\\\

[5] G. L. Murphy,   \textit{Phys. Rev. D} \textbf{8} 4231 (1973).\\\

[6] R. Garc´ýa-Salcedo and N. Breton, \textit{IJMPA} \textbf{15} 4341 (2000).\\\

[7] R. Garcýa-Salcedo and N. Breton, {\it Class. Quant. Grav.}
{\bf 22} 4783 (2005).\\

[8] M. Novello, {\it Int. J. Mod. Phys. A} {\bf 20} 2421 (2005).\\

[9] V. A. De Lorenci, R. Klippert, M. Novello, and J. M. Salim,
{\it Phys. Rev. D} {\bf 65} 063501 (2002).\\

[10] M. Novello, L. A. R. Oliveira and J. M. Salim, {\it Class.
Quantum Grav.} {\bf 13} 1089 (1996); G. Mun$\tilde{o}$z, {\it Am.
J. Phys.} {\bf 64} 1285 (1996).\\

[11] W. Heisenberg and H. Euler, {\it Z. Phys.} {\bf 98} 714
(1936); J. Schwinger, {\it Phys. Rev.} {\bf 82} 664 (1951).\\

[12] H. Salazar, A. Garcia and J. Pleba$\acute{n}$ski,  \textit{J.
Math. Phys.} \textbf{28} 2171 (1987); E. Ayo$\acute{n}$-Beato and
A. Garcia,
\textit{Phys. Rev. Lett.} \textbf{80} 5056 (1998).\\\

$[13]$ T. Jacobson, {\it Phys. Rev. Lett.} {\bf 75} 1260
(1995).\\

$[14]$ J. D. Bekenstein, {\it Phys. Rev. D} {\bf 7} 2333 (1973);
S. W. Hawking, {\it Commun. Math. Phys.} {\bf 43} 199 (1975); J.
M. Bardeen, B. Carter and S. W. Hawking, {\it Commun. Math. Phys.}
{\bf 31} 161 (1973).\\

$[15]$ E. Verlinde, hep-th/0008140.\\

$[16]$ G. W. Gibbons and S. W. Hawking, {\it Phys. Rev. D} {\bf 15} 2738 (1977).\\

$[17]$ R. G. Cai and S. P. Kim, {\it JHEP} {\bf 02} 050 (2005); Y.
Gong and A. Wang, {\it Phys. Rev. Lett.} {\bf 99} 211301 (2007);
R-G.  Cai and L-M. Cao, {\it Phys. Rev. D} {\bf 75} 064008 (2007);
X-H. Ge,  {\it Phys. Lett. B} {\bf 651} 49 (2007); 6.  M. Akbar
and R-G.  Cai, {\it Phys. Lett. B} {\bf 635} 7 (2006).\\

$[18]$ B. Wang, Y. G. Gong and E. Abdalla, {\it Phys. Rev. D} {\bf 74} 083520 (2006).\\

$[19]$ Y. Gong, B. Wang and A. Wang, {\it JCAP} {\bf 01} 024 (2007).\\

$[20]$ L. N. Granda and A. Oliveros, {\it Phys. Lett. B} {\bf 669}
275 (2008); Y. Gong, B. Wang and A. Wang, {\it JCAP} {\bf 01} 024
(2007); T. Padmanabhan, {\it Class. Quantum Grav.} {\bf 19} 5387
(2002); R. -G. Cai and N. Ohta, {\it Phys. Rev. D} {\bf 81}
084061; R. G. Cai and L. -M. Cao, {\it Nucl. Phys. B} {\bf 785}
135 (2007); M. Akbar and R. -G. Cai, {\it Phys. Lett. B} {\bf 635}
7 (2006); R. -G. Cai, L. -M. Cao, Y. -P. Hu and S. P. Kim, {\it
Phys. Rev. D} {\bf 78} 124012 (2008).\\

$[21]$ M. R. Setare and S. Shafei, {\it JCAP} {\bf 09} 011
(2006).\\

$[22]$ M. R. Setare, {\it JCAP} {\bf 01} 023 (2007).\\

$[23]$ M. R. Setare, {\it Phys. Lett. B} {\bf 641} 130 (2006).\\

$[24]$ G. Izquierdo and D. Pavon, {\it Phys. Lett. B} {\bf 633}
420 (2006).\\

$[25]$ S. Bhattacharya and U. Debnath, arXiv:1006.2600 [gr-qc]; A.
Das, S. Chattopadhyay and U. Debnath, arXiv:1104.2378 [physics.gen-ph].\\

$[26]$ C. S. Camara, M. R. de Garcia Maia, J. C. Carvalho and J.
A. S. Lima, {\it Phys. Rev. D} {\bf 69} 123504 (2004).\\

$[27]$  R. C. Tolman and P. Ehrenfest, \textit{Phys. Rev.}
\textbf{36}, 1791 (1930); M. Hindmarsh and A. Everett,
\textit{Phys. Rev. D} \textbf{58}, 103505 (1998).
\\

$[28]$ R. Bousso, {\it Phys. Rev. D} {\bf 71} 064024 (2005).\\

\end{document}